\documentclass[aps,prb,groupedaddress,twocolumn]{revtex4}

\usepackage{graphicx}

\begin{document}

\title{Two-way shape memory effect and mechanical properties of
Pulse Discharge Sintered Ni$_{2.18}$Mn$_{0.82}$Ga}

\author{T.~Takagi}
\author{V.~Khovailo}
\author{T.~Nagatomo}
\affiliation{Institute of Fluid Science, Tohoku University, Sendai
980--8577, Japan}

\author{M.~Matsumoto}
\author{M.~Ohtsuka}
\affiliation{Institute of Multidisciplinary Research for Advanced
Materials, Tohoku University, Sendai 980-8577, Japan}

\author{T.~Abe}
\affiliation{National Institute of Advanced Industrial Science and
Technology, Tohoku Center, Sendai 983--8551, Japan}

\author{H.~Miki}
\affiliation{Faculty of Systems, Science and Technology, Akita
Prefectural University, Honjo 015-0055, Japan}

\begin{abstract}
Mechanical and shape memory properties of a polycrystalline
Ni$_{2.18}$Mn$_{0.82}$Ga alloy prepared by a PDS (Pulse Discharge
Sintering) method were investigated. It was found that the
material demonstrates the two-way shape memory effect after a
loading - unloading cycle performed in the martensitic state,
i.~e. essentially without special training. The samples exhibiting
the two-way shape memory effect show a significant enhancement in
the magnitude of magnetic-field-induced strain.
\end{abstract}

\maketitle

\section{Introduction}

Beginning from 1996 when Ullakko and co-worker~\cite{1-u} reported
on a magnetic-field-induced strain of about 0.2\% in a
non-stoichiometric Ni-Mn-Ga single crystal, research in this field
has attracted a considerable attention due to the great
technological potential of this effect. Later this phenomenon was
found in other compounds exhibiting shape memory effect when in
the ferromagnetic state, such as Fe-Pd and Fe-Pt.~\cite{2-j,3-k}
Whereas Fe-Pt and Fe-Pd show magnetic-field-induced strains not
exceeding 1\%, in Ni-Mn-Ga single crystals the values of
magnetic-field-induced strains can be as large as 6\% (Ref.~4).
The mechanism of this phenomenon is believed to be the
redistribution of different twin variants under action of a
magnetic field.~\cite{5-o} Since these giant strains are attained
in comparatively low magnetic fields (less then 1.5~T), they are
easy suppressed by an external stress. For instance, the 6\%
field-induced strain was completely blocked by a compressive
stress of order 2 MPa.~\cite{4-m} This suggests that the giant
magnetostrains arising from the process of twin-boundary motion
might be useful for large stroke and small force applications.

Another way to attain a large magnetic-field-induced strain in the
ferromagnetic shape memory alloys is a shift of the martensitic
transition temperature caused by a magnetic field.~\cite{6-v,7-d}
However, in this case the magnitude of the applied field must be
high in order to overcome the temperature hysteresis of
martensitic transformation. From general consideration, it can be
expected that the maximum achievable strain in this case is equal
to the striction of the transition or even more if material is
trained for the two-way shape memory effect. The work output has
to be large (as in conventional shape memory alloys), which can be
useful for mediate stroke and large force applications.

From the point of view of widespread use of magnetic-field-induced
strains, observed in the ferromagnetic shape memory alloys, there
is a need to investigate polycrystalline materials. Previous
studies of a series of Ni$_{2+x}$Mn$_{1-x}$Ga ($x = 0.16 - 0.20$)
showed that, among the composition studied,
Ni$_{2.18}$Mn$_{0.82}$Ga is characterized by a considerable
striction of transition.~\cite{8-t} Because of that we have
studied Ni$_{2.18}$Mn$_{0.82}$Ga prepared by a Pulse Discharge
Sintering (PDS) process.

\section{Samples preparation and measurements}

Ingots of the Ni$_{2.18}$Mn$_{0.82}$Ga composition were prepared
by arc-melting of high-purity initial elements. The ingots were
annealed at 1100~K for 9~days and quenched in ice water.  A part
of the arc-melted ingots was used to fabricate PDS samples. For
this aim the arc-melted ingots initially were crushed into
particles and ground into fine powder with a particle size less
than 53~$\mu$m. Meshed powder was filled in a graphite die with
two graphite punches. The die was set in a pulse discharge system
(Sodick Co., Ltd). The pulse discharge system was evacuated to a
vacuum of 3~Pa prior to the sintering process. Maximum pressure
and temperature during the PDS process were equal to 80~MPa and
1173~K, respectively. Disc-shaped billets with thickness of 6~mm
were sintered. X-ray diffraction measurements of the samples,
performed in a wide temperature range, showed that the
high-temperature austenitic structure has a cubic modification
whereas the low-temperature martensitic phase has a complex
tetragonally based crystal structure. Samples with dimensions of
$3\times 3\times 6$ mm$^3$ were spark-cut from the billets.
Temperature and magnetic field dependencies of strain were
measured by a strain gage technique. For this aim a non-magnetic
strain gage with a compensated temperature range from 273 to 423~K
was attached along the longest dimension of the samples. The
configuration of the experimental setup allowed detecting the
relative change in the length of a specimen with an accuracy of
0.005\%. The samples were inserted into a variable temperature
chamber of a superconducting magnet. Temperature was monitored by
a Lake Shore calibrated platinum resistance thermometer with an
accuracy of 0.1~K. Stress-strain measurements and compression of
samples were done at room temperature by an Instron machine.

\begin{figure}[t]
\begin{center}
\includegraphics[width=\columnwidth]{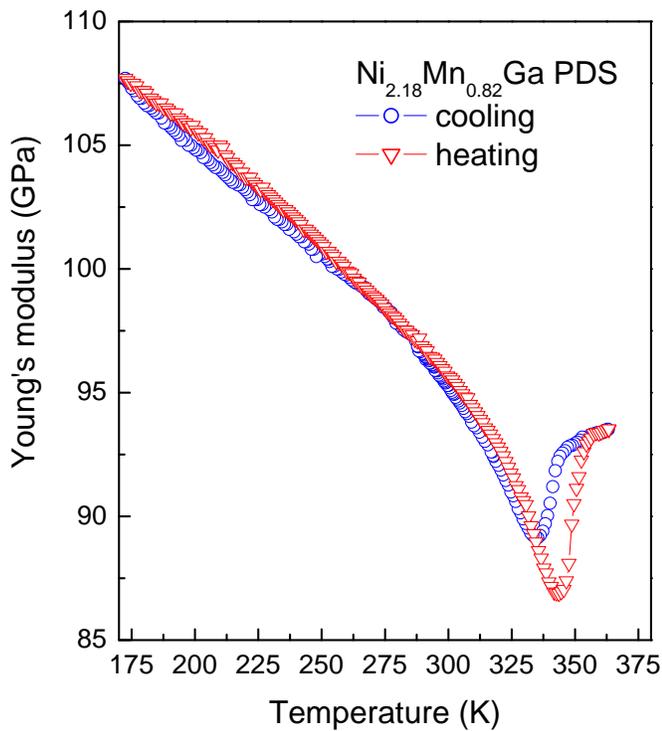}
\caption{Temperature dependencies of Young's modulus of a
Ni$_{2.18}$Mn$_{0.82}$Ga PDS sample.}
\end{center}
\end{figure}

\section{Results and discussion}

Shown in Fig.~1 are the temperature dependencies of Young's
modulus of a Ni$_{2.18}$Mn$_{0.82}$Ga PDS sample during heating
and cooling. These dependencies were obtained from ultrasonic
measurements performed in a temperature interval from 173 to
373~K. The value of Young's modulus was evaluated through the
formula

$$E = \rho \frac{v_s^2(3v_l^2 - 4v_s^2)}{(v_l^2 - v_s^2)},$$

\noindent where $\rho$, $v_l$ and $v_s$ are the density of the
material and the velocities of longitudinal and shear waves,
respectively. Marked dips on the temperature dependencies of
Young's modulus correspond to the direct and reverse martensitic
transformations. As evident from Fig.~1, the Young's modulus of
Ni$_{2.18}$Mn$_{0.82}$Ga at room temperature is equal to 95~GPa.

The results of compression tests for Ni$_{2.18}$Mn$_{0.82}$Ga
prepared by the PDS method and by a conventional arc-melting
technique are shown in Fig.~2. The compression tests were done at
room temperature with the same speed of compression for both the
samples. The comparison of these curves clearly indicates that the
PDS sample shows higher yield strength than the arc-melted one.
This characteristic is of importance for some practical
applications, and it makes the PDS materials more attractive in
this sense.

\begin{figure}[t]
\begin{center}
\includegraphics[width=\columnwidth]{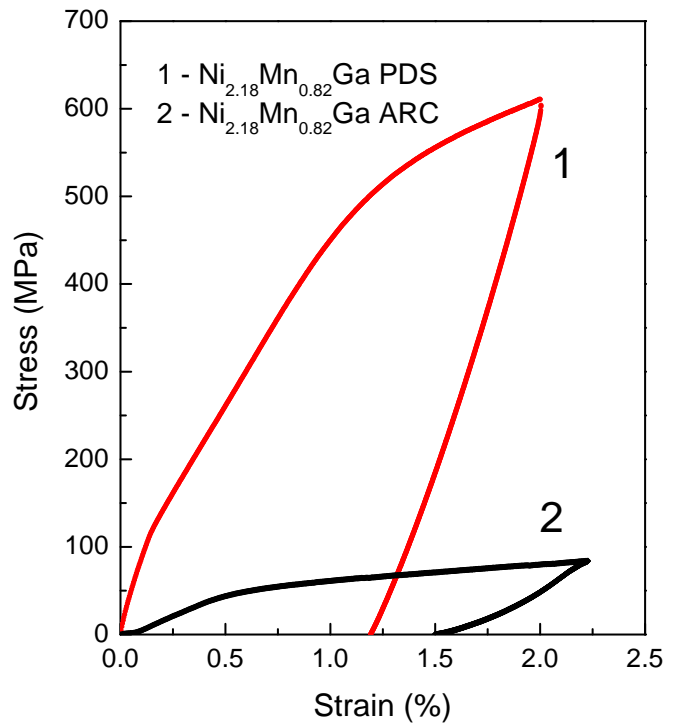}
\caption{Curves of compression test for a Ni$_{2.18}$Mn$_{0.82}$Ga
PDS sample (curve~1) and a Ni$_{2.18}$Mn$_{0.82}$Ga arc-melted
sample (curve~2).}
\end{center}
\end{figure}

\begin{figure}[b]
\begin{center}
\includegraphics[width=\columnwidth]{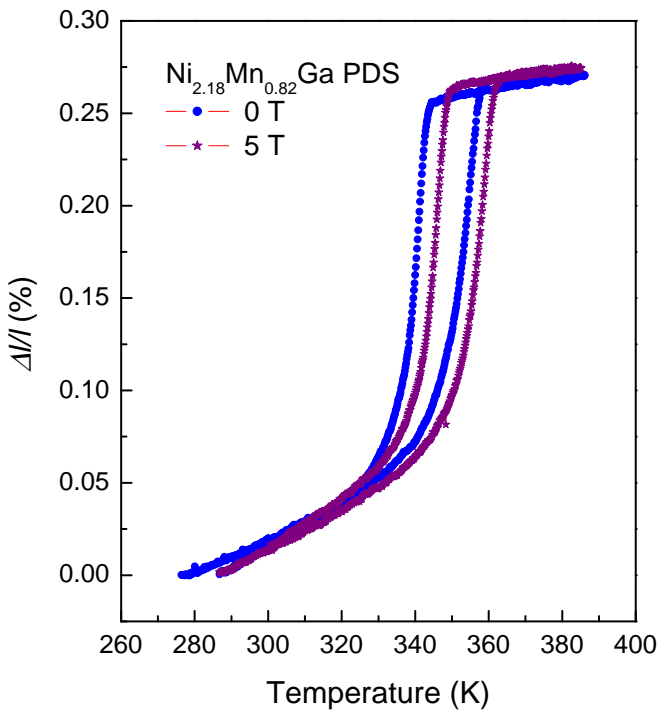}
\caption{Temperature dependencies of strain in a
Ni$_{2.18}$Mn$_{0.82}$Ga PDS sample measured in zero and 5~T
magnetic fields.}
\end{center}
\end{figure}

\begin{figure}[t]
\begin{center}
\includegraphics[width=\columnwidth]{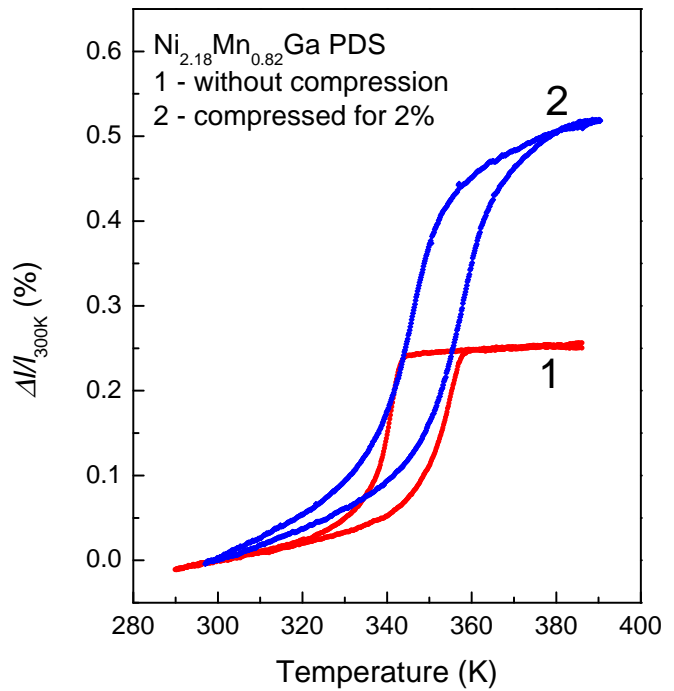}
\caption{Compression-induced two-way shape memory effect in a
Ni$_{2.18}$Mn$_{0.82}$Ga PDS sample (curve~2). Curve~1 shows the
temperature dependencies of strain in a stress-free
Ni$_{2.18}$Mn$_{0.82}$Ga PDS sample.}
\end{center}
\end{figure}

Figure~3 shows the temperature dependencies of strain measured in
a Ni$_{2.18}$Mn$_{0.82}$Ga PDS sample upon cooling and heating in
zero and 5~T magnetic fields. In zero magnetic field the sample
length monotonously increases upon heating up to the onset of the
reverse martensitic transformation, $A_s = 343$~K. The martensite
- austenite transformation is accompanied by a rapid increase in
the sample length, which flattens out at austenite finish
temperature $A_f = 357$~K. Subsequent cooling down results in the
direct martensitic transformation at $M_s = 342$~K, which is
accompanied by a shortening of the sample. As evident from Fig.~3,
the striction of the transition is approximately 0.18\%. The
temperature dependencies of strain measured in a magnetic field of
5~T revealed that the striction of the transition remains
essentially the same as in the case of the measurements without
magnetic field. The results of the measurements performed in zero
and 5~T magnetic fields leads to the conclusion that the
application of the magnetic field results in an upward shift of
the characteristic temperatures of the martensitic transition with
a rate of about 1~K/T. This value agrees very well with the
results reported for polycrystalline Ni$_{2+x}$Mn$_{1-x}$Ga
prepared by arc-melting method.~\cite{7-d}

An interesting feature of the PDS samples is that the two-way
shape memory effect can be induced in these materials by a simple
loading - unloading cycle. This feature is presented in Fig.~4.
Indeed, the increase in the sample length caused by the
martensitic transition is approximately 0.18\% in the case of the
sample which was not subjected to compression. Another sample cut
from the same ingot was compressed for 2\% at room temperature in
the martensitic state. After unloading the residual deformation
was approximately 1.2\% (Fig.~2). The sample recovered
approximately 75\% of its initial length upon the first heating,
showing shape memory effect. The subsequent cooling - heating
process revealed that for this sample the change in the length
associated with the martensitic transformation increased twofold
as compared with the uncompressed sample and reached 0.4\%. It is
also seen from Fig.~4 that the change in slope of the curves at
the characteristic temperatures of martensitic transformation in
the compressed sample becomes less pronounced than in the
compression-free sample. Further themocyclings demonstrated that
this compression-induced two-way shape memory effect does not
degrade and the 0.4\% change in the length of the sample is very
well reproducible at least up to the tenth heating - cooling
cycle. It is interesting to note that a well-defined two-way shape
memory effect has also been found recently in Ni-Mn-Ga thin
films.~\cite{9-o}

To study this compression-induced two-way shape memory effect in
more detail, several Ni$_{2.18}$Mn$_{0.82}$Ga PDS samples were
compressed for values of strain, ranging from 1 to 6\%. After
unloading the residual deformation in these samples was from 0.4
to 4\%, respectively. The behavior of the samples upon the first
heating process was found to be dependent on the value of residual
deformation. Thus, the sample with 0.4\% residual deformation
showed a perfect shape memory effect whereas the sample with the
largest residual deformation did not exhibit shape memory effect,
which means that the residual deformation in this sample was
essentially plastic one. The two-way shape memory effect was found
only in the samples compressed for 2 and 3\%. Together with the
observation that the sample compressed for 2 and 3\% did not
revert the original shape completely after the first heating,
these facts evidence that the two-way shape memory effect appears
when the applied stress exceeds some critical limit which is
enough for the occurrence of an irreversible slip. We suggest that
it arises from the strain field of dislocations induced upon
compression.

\begin{figure}[h]
\begin{center}
\includegraphics[width=\columnwidth]{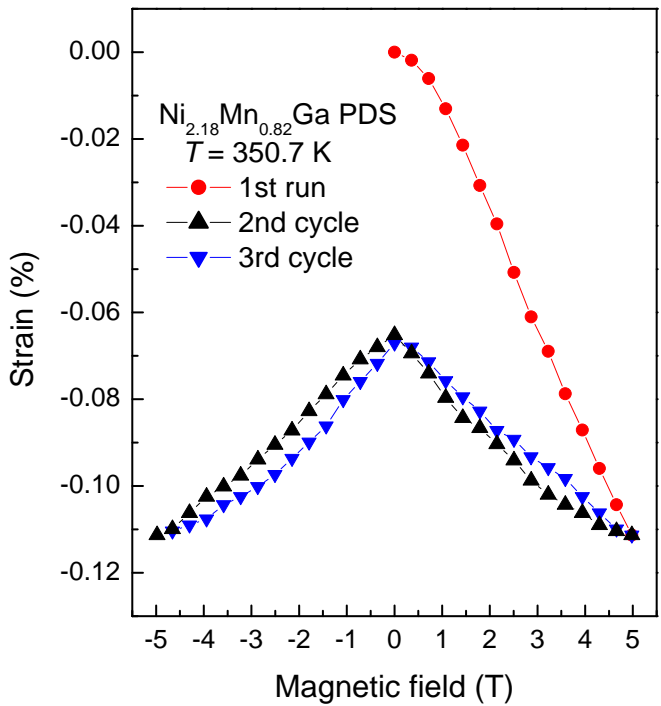}
\caption{Strain induced by a magnetic field in the temperature
interval of the direct martensitic transformation in a
Ni$_{2.18}$Mn$_{0.82}$Ga sample, exhibiting two-way shape memory
effect.}
\end{center}
\end{figure}

The measurements of magnetic-field-induced strain in the
temperature interval of the martensitic transformation in the
PDS-sintered Ni$_{2.18}$Mn$_{0.82}$Ga samples showed that these
samples exhibit rather small values of magnetic-field-induced
strain even in high (up to 5~T) magnetic fields. For instance, in
the compression-free Ni$_{2.18}$Mn$_{0.82}$Ga sample this value is
equal to 0.02\% in a 5~T magnetic field. However, in the case of
the samples which demonstrate the compression-induced two-way
shape memory effect the magnetic-field-induced strain is six times
greater as compared to the samples without two-way shape memory
effect. In fact, such a tendency could be expected, since in this
case the magnetic-field-induced strain is due to the shift of the
martensitic transition temperature caused by the applied magnetic
field. The magnetic-field-induced strain is proportional to the
relative change in the dimension of the sample per 1~K and this
characteristic is much better in the samples with the two-way
shape memory effect. It can be suggested that an appropriate
training procedure of the PDS material will results in enhancement
of the magnitude of the two-way shape memory effect, leading to an
increase in the value of magnetic-field-induced strain. It should
be noted, however, that the strain of about 0.12\% is not
perfectly recovered. Figure 5 shows that the reversible
magnetic-field-induced strain is equal to 0.06\%. This value of
magnetic-field-induced strain has been reversible for many cycles
of application and removal of the magnetic field.

\section{Conclusion}

In conclusion, the most interesting findings of this study are
that the two-way shape memory effect can be induced in the
Ni$_{2.18}$Mn$_{0.82}$Ga PDS materials by an ordinary compression
of the samples in the martensitic state. The samples with the
two-way shape memory effect show a significant enhancement in the
magnitude of magnetic-field-induced strain observed in the
temperature interval of martensitic transformation.

\section*{Acknowledgements}

This work was partially supported by the Grant-in-Aid for
Scientific Research (C) No.~11695038 from the Japan Society of the
Promotion of Science.


\begin{thebibliography}{40}

\bibitem{1-u} K.~Ullakko, J.K.~Huang, C.~Kantner, R.C.~O'Handley, and
V.V.~Kokorin, Appl. Phys. Lett. \textbf{69}, 1966(1996).

\bibitem{2-j} R.D.~James and M.~Wuttig, Philos. Mag. A \textbf{77}, 1273 (1998).

\bibitem{3-k} T.~Kakeshita, T.~Takeuchi, T.~Fukuda,
M.~Tsujiguchi, T.~Saburi, R.~Oshima, and S.~Muto, Appl. Phys.
Lett. \textbf{77}, 1502 (2000).

\bibitem{4-m} S.J.~Murray, M.~Marioni, S.M.~Allen, R.C.~O'Handley,
and T.A.~Lograsso, Appl. Phys. Let. \textbf{77}, 886 (2000).

\bibitem{5-o} R.C.~O'Handley, J. Appl. Phys. \textbf{83}, 3263 (1998).

\bibitem{6-v} A.N.~Vasil'ev, A.D.~Bozhko, V.V.~Khovailo, I.E.~Dikshtein,
V.G.~Shavrov, V.D.~Buchel'nikov, M.~Matsumoto, S.~Suzuki,
T.~Takagi, and J.~Tani, Phys. Rev. B \textbf{59}, 1113 (1999).

\bibitem{7-d} I.E.~Dikshtein, D.I.~Ermakov, V.V.~Koledov, L.V.~Koledov,
T.~Takagi, A.A.~Tulaikova, A.A.~Cherechukin, and V.G.~Shavrov,
JETP Lett. \textbf{72}, 373 (2000).

\bibitem{8-t} T.~Takagi, V.~Khovailo, T.~Nagatomo, H.~Miki, M.~Matsumoto,
T.~Abe, Z.~Wang, E.~Estrin, A.~Vasil'ev, and A.~Bozhko, Trans.
Mater. Res. Soc. Jpn. \textbf{26}, 197 (2001).

\bibitem{9-o} M.~Ohtsuka and K.~Itagaki, Int.
J. Appl. Electromagn. Mech. \textbf{12}, 49 (2000).



\end{thebibliography}
\end{document}